\documentclass[letterpaper,10pt]{article} 
\usepackage{osameet3} %% use version 3 for proper copyright statement

\usepackage{amsmath,amssymb}
\usepackage{upgreek}
\usepackage{multirow}
\usepackage{pgfplots}
\usepackage{graphicx}
\usepackage{subcaption}
\usepackage{mwe}
\usepackage{adjustbox}
\usepackage{xcolor}
\usepackage{collcell}
\usepackage{multirow}
\usepackage{mathtools}
\usepackage[colorlinks=true,bookmarks=false,citecolor=blue,urlcolor=blue]{hyperref} %pdflatex
\usetikzlibrary{arrows, arrows.meta}

\begin{document}

\title{Domain Adaptation: the Key Enabler of Neural Network  Equalizers in Coherent Optical Systems}
% \title{Neural Network Equalizers for Coherent Optical Systems: Bridging the Reality Gap}
% \title{Sim-to-Exp: Bridging the Reality Gap in Neural Network Equalizers for Coherent Optical Systems}
\author{Pedro J. Freire\textsuperscript{(1)}, Bernhard Spinnler\textsuperscript{(2)}, Daniel Abode\textsuperscript{(1)}, Jaroslaw E. Prilepsky \textsuperscript{(1)}, Abdallah A. I. Ali \textsuperscript{(1)}, Nelson Costa\textsuperscript{(3)}, Wolfgang Schairer\textsuperscript{(2)}, Antonio Napoli\textsuperscript{(4)}, Andrew D. Ellis \textsuperscript{(1)}, Sergei K. Turitsyn\textsuperscript{(1)}}
\address{\textsuperscript{(1)}~Aston University, Birmingham, United Kingdom;
   \textsuperscript{(2)}~Infinera, Munich, Germany; \textsuperscript{(3)}  Infinera Unipessoal, Carnaxide, Portugal; \textsuperscript{(4)}~Infinera, London, United Kingdom
  
   }
  \vspace{-0.5mm}
\email{p.freiredecarvalhosourza@aston.ac.uk}

%% Uncomment the following line to override copyright year from the default current year.
\copyrightyear{2021}

% \begin{abstract}
% We propose a convolutional-recurrent channel equalizer and experimentally demonstrate 1 dB Q-factor improvement both in a single-channel and a 96 x WDM, DP-16QAM transmission over 450 km of TWC fiber. The proposed equalizer outperforms previous NN-based approaches and the 3-step-per-span DBP.
% \end{abstract}

\begin{abstract}
We introduce the domain adaptation and randomization approach for calibrating neural network-based equalizers for real transmissions, using synthetic data. The approach renders up to 99\% training process reduction, which we demonstrate in three experimental setups.
\end{abstract}

% \begin{abstract}
% We analyze the performance of several previously proposed neural network-based equalizers using the experimental data from 450~km TWC-fiber 200G DP-16QAM 33.4GBd single channel and WDM transmission runs. A new combined convolutional-recurrent neural network equalizer is introduced, with which we demonstrate a 1.35~dB Q-factor improvement in a 96 channel WDM scenario, beating the results rendered by other deep neural network-based equalizers.\redA{too long}
% \end{abstract}

% \vspace{-1.5mm}

% \huge

\section{Introduction}
\vspace{-1.5mm}
The use of neural networks (NNs) to compensate for optical transmission impairments has gained considerable momentum thanks to NNs' remarkable capability to mimic the inverse transfer function of nonlinear optical channels with, simultaneously, a high noise tolerance~\cite{Pedro2021,Pedro2020,Koike_Akino2020,Bertold2020}. The three cornerstones for the efficient NNs' functioning in channel equalization are: good and reliable performance, acceptable computational complexity, and configuration flexibility. Regarding the latter, unfortunately, the straightforward adaptation of NN-based solutions for the efficient NN's operation typically requires large training datasets corresponding to each peculiar scenario change~\cite{Pedro2021,Azzimonti}. However, in realistic conditions, the collection of very large datasets is typically inhibited by practical limitations. One of the most critical problems related to the NN's flexibility is the ``reality gap'': the discrepancy between reality and numerical modeling that prevents the simulated NN to be applied directly to a real-world system. So far, the reality gap problem has been addressed in robotics, image denoising, and natural language processing~\cite{Tobin2017,Yoonsik}. In optical communication systems, this problem applies, e.g., to channel equalization, where the simulated channel transfer function provides only an approximation of the more complex real transmission system, such that we need a considerable amount of real data to retrain the NNs to achieve the desired performance. In contrast, numerical simulation data is often abundantly available for a wide range of different transmission scenarios, so it is tempting to use that synthetic data to enable the NN's functioning in realistic conditions. 
% \redA{concerning "abundantly available", i think that synthetic data also help as you can keep generating them and store in the cloud for a rapid usage whenever you need them.} % The most common techniques used to bridge reality and simulations are transfer learning and domain randomization.

In this work, we propose a method for adapting the NN-based equalizers to changes in the transmission scenario. Our calibration method is based on models pre-trained with data obtained from numerical simulation and requires 99\% fewer calibration resources (epochs and real training data from experiment) to become fully functional in a real transmission system. The method's performance is demonstrated in three different experimental setups.

\vspace{-1.5mm}

\section{NN-based Equalizer Calibration Method}
\begin{figure}[htbp]
\centering
\begin{minipage}{.7\textwidth}
  \centering
  \includegraphics[width=\linewidth]{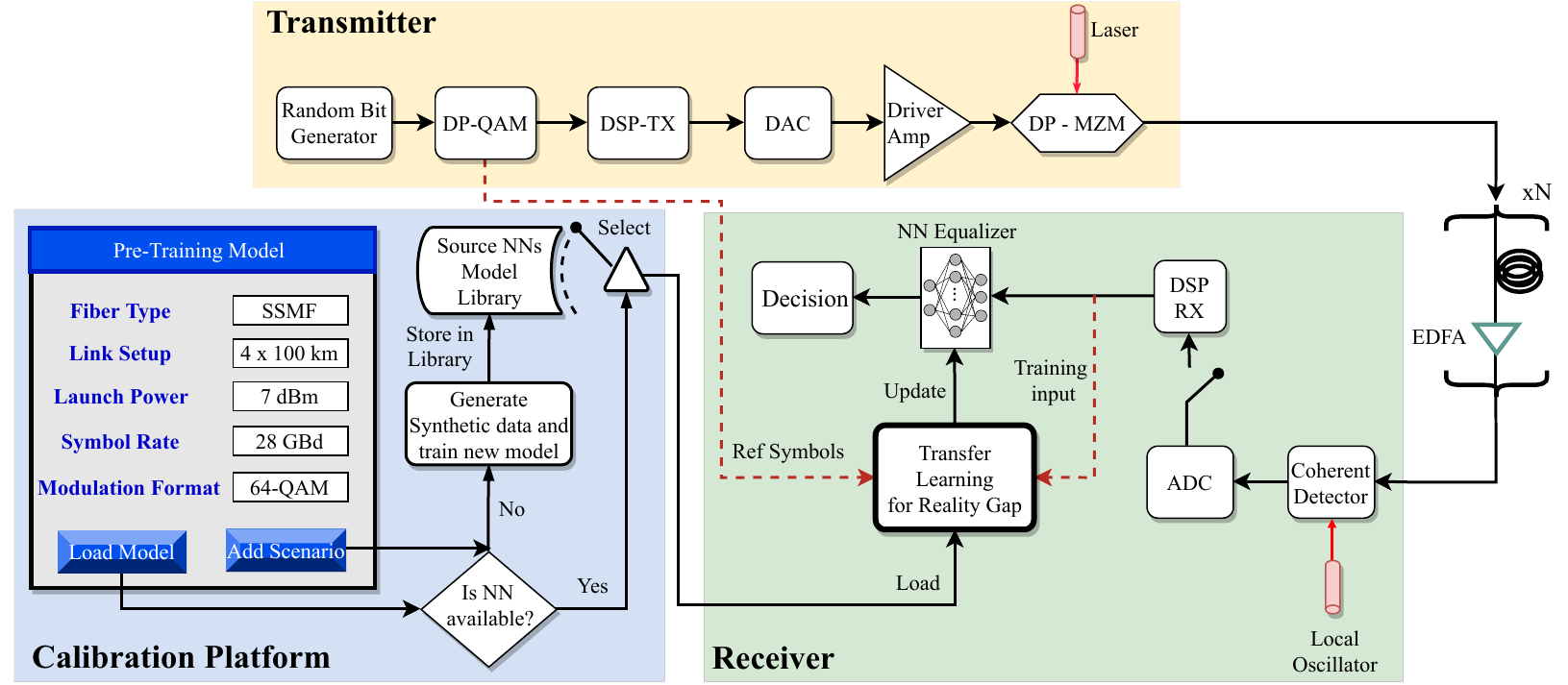}
  \captionof{figure}{\footnotesize Experimental setup and proposed integrated method for bridging the reality gap using a graphical user interface (GUI).}
  \label{fig:setup}
\end{minipage}%
\hspace{1mm}
\begin{minipage}{.25\textwidth}
  \centering
  \includegraphics[width=1.2\linewidth]{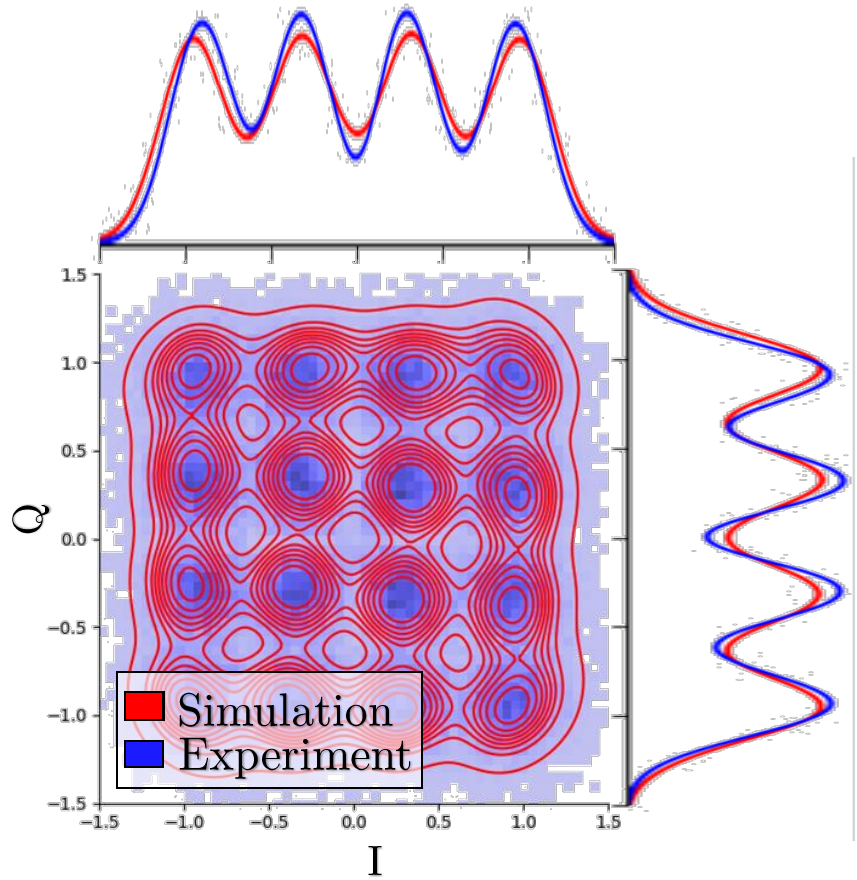}
  \captionof{figure}{\footnotesize PDF of simulation and experimental datasets of the setup described in Fig.3 B.}
  \label{fig:distribution}
\end{minipage}
\end{figure}

\vspace{-1.5mm}
Our calibration method consists of two steps: the \textit{domain randomization} using synthetic data to pre-train the model, and the \textit{transfer learning} (TL), a parameter-based \textit{domain adaption} method used to fine-tune the NN equalizer to the real transmission system.
Domain randomization is a systematic approach for data generation that aims at improving the generalization of machine learning algorithms to new environments. The domain randomization is carried out by generating training data from a random distribution with given desired properties and storing it into a library which can be assessed by the NN. This is a critical step because the optical fiber parameters considered in the numerical simulation will most likely not perfectly match those from the actually deployed fiber.
In practical terms, and as shown in Fig.\ref{fig:setup}, the calibration method can proceed in two possible ways: i) if a set of numerically generated data -- for the transmission system under evaluation -- already exists in our library then the corresponding pre-trained NN model is loaded and sent to the TL step; or ii) if the existing numerical data is far from the actual transmission system, the method will create new synthetic data and train the NN model using this new dataset. This pre-trained NN model is saved in the library and sent to the next TL step. Please note that the domain randomization takes place in step ii): we generate twenty datasets with $2^{19}$ symbols where, for each numerical simulation run, the dispersion coefficient $D$, the effective nonlinearity coefficient $\gamma$, and the fiber loss coefficient $\alpha$ are picked randomly from within a pre-defined range\footnote{The NN cannot generalize for noticeably different distributions, so the range of parameters cannot be too wide. The ranges in this work have been selected empirically and worked well for all cases considered. }. When using SSMF, the ranges of the fiber parameters are: $\alpha = [0.19 -0.22]$ dB/km, $D = [16.5-17.5]$ ps/(nm$\cdot$km), and $\gamma = [1.1-1.5]$ (W$\cdot$km)$^{-1}$. When using TWC fiber, the ranges are: $\alpha = [0.2 -0.25]$ dB/km, $D = [2.5 - 3.5]$ ps/(nm$\cdot$km), and  $\gamma = [2 - 3]$~(W$\cdot$km)$^{-1}$. The simulator used for generating data is described in Ref.\cite{Pedro2021,Pedro2021_transfer}. Then, for each NN training epoch, we randomly select  $2^{18}$ symbols from one of 20 generated synthetic datasets (corresponding to the desired transmission setup) to use in the NN model's pre-training. A similar approach was used with great success in robotics \cite{Tobin2017}. The probability distribution functions (PDFs) of both simulated and experimental data after transmission of a dual-polarization (DP) 34.4 GBd 16-QAM single-channel launched with an optical power of 9 dBm along a 5$\times$50 km SSMF link are plotted in Fig.~\ref{fig:distribution}. The analysis of Fig.~\ref{fig:distribution} shows that there are differences between the two PDFs that still need to be learned for efficient optical channel equalization. This is tackled by utilizing TL.

\vspace{-3.5mm}

The TL is based on the fact that a suitable target estimate (experiment) may be created by adjusting the parameters of a pre-trained source estimator (simulation) using a small set of labeled target data. The method entails fitting a NN to the target data using an objective function (e.g. the mean square error (MSE) loss), but with its weights initialized based on the source data. In \cite{Pedro2021_transfer}, we showed how the TL can be used to reduce training resources for NN equalizers when changing the optical power, modulation format, and symbol rate using simulation data.

In this paper, we apply the same methodology to improve the learning process from synthetic to experimental data. We used the same equalizer architecture as in \cite[Fig.~1]{Pedro2021_transfer}, the CNN+biLSTM equalizer, with 224 filters, kernel size 10, 226 hidden units, and 25 input taps. We found that, by freezing the CNN layers and just retraining the LSTM part, we can achieve a simple and fast calibration of the NN. For the considered CNN+biLSTM NN, we incorporated the MSE estimator and the classical Adam algorithm for the stochastic optimization step with the learning rate set to 0.001. The training was carried out for up to 200 epochs with a batch size of 1000, which has proven to be large enough for the convergence of the NN model\footnote{For the pre-training with synthetic data, we used the same hyperparameters as in the TL step, but considered 1000 epochs instead.}. Additionally, the total experimental dataset used was composed of $2^{18}$ symbols for the training dataset and $2^{18}$ independently generated symbols for the testing phase, with different random seeds, which guarantees a cross-correlation below 0.03 between the training and testing datasets. All precautions against possible overfitting and overestimation were taken~\cite{Eriksson}. 
 
\vspace{-1.5mm}

\section{Results and Discussions}
\vspace{-1.5mm}
Fig.~\ref{fig:setup} shows the general setup used in our three experiments (A, B, and C) performed in two different labs: the UK lab (A) and the German Lab (B and C). At the transmitter, a single channel DP-(A: 64-QAM; B/C: 16-QAM) (A: 28; B/C: 34.4)~GBd symbol sequence was mapped out of data bits generated by a $2^{32} - 1$ order PRBS. A digital RRC filter with roll-off 0.1 was applied to limit the channel bandwidth to 37.5~GHz. The filtered digital samples were uploaded to a digital-to-analog converter (DAC) operating at 88 GSamples/s. The outputs of the DAC were amplified by a four-channel electrical amplifier that drove a dual-polarization I/Q Mach–Zehnder modulator, modulating the continuous waveform carrier produced by an external cavity laser at $\lambda = 1.55 \mu m$. The optical launch powers used in this paper are (A: 7 dBm; B: 9 dBm; C: 3 dBm). These numbers guarantee a highly nonlinear transmission regime.
The resulting optical signal was then transmitted along (A: 4$\times$100 km; B: 5$\times$50 km; C: 9$\times$50 km) spans of (A/B: SSMF; C: TWC) optical fiber with EDFA amplification. At the RX side, the optical signal was converted into the electrical domain using an integrated coherent receiver. The resulting signal was sampled at 50~Gsamples/s by a digital sampling oscilloscope and processed by an offline DSP which includes digital chromatic dispersion compensation, multiple-input-multiple-output (MIMO) equalization, clock recovery, and pilot-aided carrier recovery.  The NN equalizer is then placed after this standard DSP and the system performance is evaluated in terms of the mutual information's (MI) lower bound \cite[Eq.~(8)]{Catuogno}.

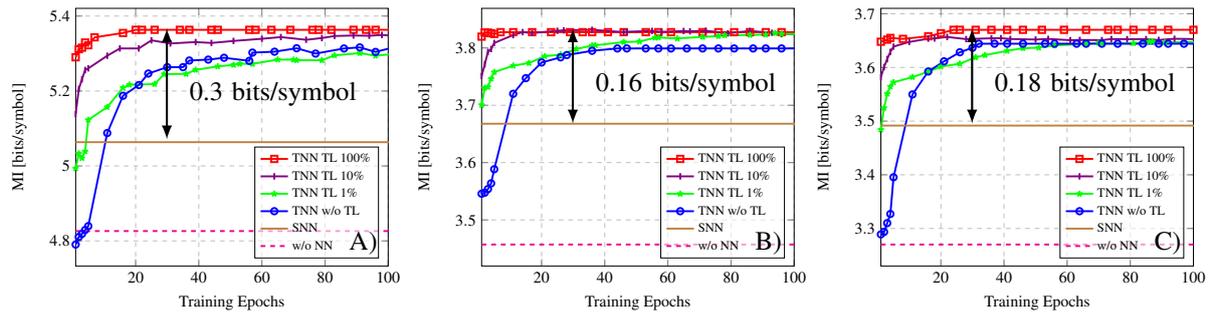
\begin{figure*}[ht!] 
  \begin{subfigure}[b]{0.33\linewidth}
    \centering
 \begin{tikzpicture}[scale=0.6]
    \begin{axis} [ylabel={BER}, 
        xlabel={Training Epochs},
        ylabel={MI [bits/symbol]},
        grid=both,  
        xmin=1, xmax=100,
    	%xtick={-6, ..., 5},
    % 	ymin=0.00001, ymax=0.1,
    %     ymode=log,
        %log ticks with fixed point,
        legend style={legend pos=south east, legend cell align=left,fill=white, fill opacity=0.6, draw opacity=1,text opacity=1},
    	grid style={dashed}]
        ]
    
            \addplot[color=red, mark=square, very thick]
  coordinates {(1,5.29)(2,5.3117)(3,5.3156)(4,5.3294)(5,5.3222)(7,5.3432)(16,5.3549)(20,5.3633)(22,5.3633)(26,5.3633)(34,5.3633)(37,5.3633)(44,5.3633)(46,5.3633)(56,5.3633)(58,5.3633)(63,5.3633)(69,5.3633)(76,5.3633)(78,5.3633)(82,5.3633)(89,5.3633)(93,5.3633)(96,5.3633)(101,5.3633)(107,5.3633)(111,5.3633)(117,5.3633)(123,5.3633)(131,5.3633)(139,5.3633)(146,5.3633)(151,5.3633)(157,5.3266)(161,5.3308)(166,5.3273)(171,5.3308)(176,5.3308)(181,5.3308)(186,5.3266)(191,5.3273)(196,5.3266)(201,5.3266)(206,5.3273)};
    \addlegendentry{\footnotesize{TNN TL 100\%}};

%     \addplot[color=black, mark=star, very thick]
%   coordinates {(1,5.1465)(2,5.2206)(3,5.3163)(4,5.2761)(5,5.3176)(11,5.4004)(11,5.4004)(19,5.4428)(25,5.44)(29,5.4541)(36,5.4622)(36,5.4622)(43,5.4664)(48,5.4693)(54,5.4512)(60,5.4984)(62,5.4961)(68,5.4653)(71,5.4652)(76,5.4614)(81,5.4541)(87,5.4813)(93,5.4762)(97,5.4732)(106,5.4596)(111,5.4757)(112,5.4596)(116,5.4732)(121,5.4614)(126,5.4614)(131,5.4732)(136,5.4614)(141,5.4732)(146,5.4614)(151,5.4614)(156,5.4614)(161,5.4732)(168,5.476)(173,5.476)(176,5.4596)(182,5.476)(187,5.4596)(191,5.4596)(197,5.476)(206,5.4596)};
%     \addlegendentry{\footnotesize{TNN TL 50\%}};
    
%     \addplot[color=green, mark=star, very thick]
%   coordinates {(1,4.9821)(2,5.1167)(3,5.2042)(4,5.2114)(5,5.2707)(9,5.3024)(15,5.3462)(19,5.3497)(22,5.3908)(29,5.4129)(32,5.4108)(41,5.4126)(46,5.4509)(46,5.4509)(54,5.4496)(59,5.4257)(65,5.4371)(67,5.459)(73,5.4631)(78,5.4757)(82,5.4849)(90,5.4686)(96,5.4417)(100,5.4821)(103,5.4687)(108,5.4691)(113,5.4742)(118,5.4783)(124,5.4531)(127,5.4526)(134,5.4691)(136,Inf)(141,5.4691)(146,5.4417)(151,5.4417)(156,5.4417)(161,5.4691)(168,5.476)(173,5.476)(176,5.4596)(182,5.476)(187,5.4687)(191,5.4687)(197,5.476)(206,5.4687)};
%     \addlegendentry{\footnotesize{TNN TL 20\%}};
    
    \addplot[color=violet, mark=|, very thick]
  coordinates {(1,5.1383)(2,5.2053)(3,5.2313)(4,5.256)(5,5.2604)(11,5.2941)(15,5.3136)(21,5.3136)(25,5.3353)(31,5.3268)(39,5.3307)(44,5.3282)(51,5.3336)(60,5.3394)(66,5.3439)(74,5.3367)(79,5.3413)(83,5.3468)(89,5.3481)(94,5.3474)(98,5.3496)(101,5.348)(110,5.354)(116,5.3471)(117,5.3507)(121,5.3489)(127,5.3431)(134,5.3566)(137,5.3566)(146,5.3566)(151,5.3566)(152,5.3599)(159,5.3486)(164,5.3562)(171,5.3636)(177,5.3566)(185,5.3512)(190,5.3662)(194,5.3602)(196,5.3662)(201,5.3602)};
    \addlegendentry{\footnotesize{TNN TL 10\% }};

%     \addplot[color=brown, mark=star, very thick]
%   coordinates {(1,4.7491)(2,4.8843)(3,4.9269)(4,4.9976)(5,5.0499)(11,5.1502)(15,5.2111)(20,5.2567)(24,5.2771)(29,5.2948)(32,5.3011)(38,5.3057)(44,5.3049)(46,5.3036)(55,5.3694)(60,5.3296)(65,5.3508)(67,5.3685)(76,5.3631)(77,5.3709)(86,5.3799)(86,5.3799)(92,5.3503)(97,5.3934)(104,5.3822)(107,5.38)(113,5.3958)(120,5.4031)(126,5.3907)(126,5.3907)(136,5.3952)(136,5.3952)(146,5.4545)(146,5.4545)(152,5.4091)(158,5.3938)(164,5.3974)(167,5.429)(175,5.4444)(181,5.3982)(184,5.4116)(192,5.3982)(201,5.4444)};
%     \addlegendentry{\footnotesize{TNN TL 5\% }};
    
    \addplot[color=green, mark=star, very thick]
  coordinates {(1,4.9936)(2,5.0335)(3,5.0209)(4,5.0391)(5,5.1228)(11,5.1572)(16,5.2087)(18,5.2167)(26,5.2185)(29,5.2452)(36,5.2457)(39,5.2565)(46,5.2656)(51,5.2698)(53,5.2726)(57,5.2723)(65,5.2839)(70,5.2835)(71,5.2823)(79,5.2821)(83,5.2951)(91,5.3011)(96,5.2941)(104,5.301)(107,5.3026)(114,5.3046)(118,5.3125)(126,5.3114)(127,5.312)(135,5.3053)(138,5.3067)(146,5.3039)(151,5.3168)(156,5.3165)(165,5.3233)(166,5.318)(175,5.3155)(177,5.3132)(185,5.3158)(186,5.3129)(196,5.3143)(205,5.318)};
    \addlegendentry{\footnotesize{TNN TL 1\%}};

    \addplot[color=blue, mark=o, very thick]   
    coordinates {(1,4.7903)(2,4.8105)(3,4.8186)(4,4.8291)(5,4.8395)(11,5.0878)(16,5.1873)(21,5.2159)(24,5.2466)(30,5.2636)(35,5.2637)(37,5.2814)(43,5.2838)(48,5.2887)(56,5.2805)(57,5.3025)(64,5.3049)(71,5.3141)(77,5.2999)(86,5.3135)(91,5.3164)(96,5.3043)(103,5.3193)(110,5.3215)(116,5.3215)(118,5.3215)(125,5.3215)(129,5.3215)(132,5.3215)(137,5.3215)(145,5.3215)(150,5.3215)(156,5.3215)(161,5.3089)(166,5.3082)(176,5.3018)(184,5.3003)(187,5.2951)(195,5.2979)(199,5.2929)(206,5.2931)};
    \addlegendentry{\footnotesize{TNN w/o TL}};
    
    \addplot[color=brown, very thick]     
     coordinates {(1,5.0635)(200,5.0635)};
     \addlegendentry{\footnotesize{SNN}};
    
     \addplot[color=magenta,dashed,very thick]     
     coordinates {(1,4.8262)(200,4.8262)};
     \addlegendentry{\footnotesize{w/o NN}};

    \end{axis}
     \node[text width=3cm] at (8.5,0.5) 
    {A)};
    \draw[>=latex, thick, <->] (2,2.8) -- (2,5.25);
    \node[text width=3cm] at (5,3.8) 
    {0.3 bits/symbol};
    \end{tikzpicture}
    % \vspace{-1mm}scale=0.6
    % \caption{7dBm  64-QAM SSMF link \\ (4x100~km) - LAB A Trial.}
    \label{fig3:a} 
    \vspace{4ex}
  \end{subfigure} 
  \begin{subfigure}[b]{0.33\linewidth}
    \centering
 \begin{tikzpicture}[scale=0.6]
    \begin{axis} [ylabel={MI [bits/symbol]}, 
        xlabel={Training Epochs},
        grid=both,
    	xmin=1, xmax=100,
    %	xtick={-6, ..., 2},
    % 	ymin=0.00001, ymax=0.1,
    %     ymode=log,
        %log ticks with fixed point,
        legend style={legend pos=south east, legend cell align=left,fill=white, fill opacity=0.6, draw opacity=1,text opacity=1},
    	grid style={dashed}]
        ]
        \addplot[color=red, mark=square, very thick]
  coordinates {(1,3.8197)(2,3.8268)(3,3.8267)(4,3.8257)(5,3.8242)(7,3.8275)(11,3.8275)(17,3.8275)(23,3.8275)(29,3.8275)(33,3.8275)(36,3.8275)(41,3.8275)(46,3.8275)(53,3.8275)(57,3.8275)(61,3.8275)(71,3.8275)(76,3.8275)(81,3.8275)(86,3.8275)(91,3.8275)(96,3.8275)(103,3.8275)(108,3.8275)(111,3.8275)(117,3.8275)(121,3.8275)(126,3.8275)(131,3.8275)(136,3.8275)(141,3.8275)(146,3.8275)(151,3.8275)};
    \addlegendentry{\footnotesize{TNN TL 100\%}};

%     \addplot[color=black, mark=star, very thick]
%   coordinates {(1,5.1465)(2,5.2206)(3,5.3163)(4,5.2761)(5,5.3176)(11,5.4004)(11,5.4004)(19,5.4428)(25,5.44)(29,5.4541)(36,5.4622)(36,5.4622)(43,5.4664)(48,5.4693)(54,5.4512)(60,5.4984)(62,5.4961)(68,5.4653)(71,5.4652)(76,5.4614)(81,5.4541)(87,5.4813)(93,5.4762)(97,5.4732)(106,5.4596)(111,5.4757)(112,5.4596)(116,5.4732)(121,5.4614)(126,5.4614)(131,5.4732)(136,5.4614)(141,5.4732)(146,5.4614)(151,5.4614)(156,5.4614)(161,5.4732)(168,5.476)(173,5.476)(176,5.4596)(182,5.476)(187,5.4596)(191,5.4596)(197,5.476)(206,5.4596)};
%     \addlegendentry{\footnotesize{TNN TL 50\%}};
    
%     \addplot[color=green, mark=star, very thick]
%   coordinates {(1,4.9821)(2,5.1167)(3,5.2042)(4,5.2114)(5,5.2707)(9,5.3024)(15,5.3462)(19,5.3497)(22,5.3908)(29,5.4129)(32,5.4108)(41,5.4126)(46,5.4509)(46,5.4509)(54,5.4496)(59,5.4257)(65,5.4371)(67,5.459)(73,5.4631)(78,5.4757)(82,5.4849)(90,5.4686)(96,5.4417)(100,5.4821)(103,5.4687)(108,5.4691)(113,5.4742)(118,5.4783)(124,5.4531)(127,5.4526)(134,5.4691)(136,Inf)(141,5.4691)(146,5.4417)(151,5.4417)(156,5.4417)(161,5.4691)(168,5.476)(173,5.476)(176,5.4596)(182,5.476)(187,5.4687)(191,5.4687)(197,5.476)(206,5.4687)};
%     \addlegendentry{\footnotesize{TNN TL 20\%}};
    
    \addplot[color=violet, mark=|, very thick]
  coordinates {(1,3.751)(2,3.7714)(3,3.7931)(4,3.798)(5,3.8087)(11,3.8224)(14,3.8276)(17,3.8259)(25,3.8299)(27,3.8307)(36,3.8316)(46,3.8276)(47,3.8282)(55,3.8297)(58,3.828)(65,3.8295)(67,3.829)(72,3.83)(79,3.8255)(86,3.8295)(93,3.8264)(96,3.8241)(104,3.8244)(109,3.8269)(112,3.8268)(117,3.8289)(124,3.8267)(129,3.8226)(136,3.8229)(140,3.8239)(142,3.8259)(150,3.8241)(151,3.8223)(161,3.8224)(163,3.8264)(166,3.8179)};
    \addlegendentry{\footnotesize{TNN TL 10\% }};

%     \addplot[color=brown, mark=star, very thick]
%   coordinates {(1,4.7491)(2,4.8843)(3,4.9269)(4,4.9976)(5,5.0499)(11,5.1502)(15,5.2111)(20,5.2567)(24,5.2771)(29,5.2948)(32,5.3011)(38,5.3057)(44,5.3049)(46,5.3036)(55,5.3694)(60,5.3296)(65,5.3508)(67,5.3685)(76,5.3631)(77,5.3709)(86,5.3799)(86,5.3799)(92,5.3503)(97,5.3934)(104,5.3822)(107,5.38)(113,5.3958)(120,5.4031)(126,5.3907)(126,5.3907)(136,5.3952)(136,5.3952)(146,5.4545)(146,5.4545)(152,5.4091)(158,5.3938)(164,5.3974)(167,5.429)(175,5.4444)(181,5.3982)(184,5.4116)(192,5.3982)(201,5.4444)};
%     \addlegendentry{\footnotesize{TNN TL 5\% }};
    
    \addplot[color=green, mark=star, very thick]
  coordinates {(1,3.7006)(2,3.7303)(3,3.7319)(4,3.7463)(5,3.7577)(11,3.769)(16,3.7737)(21,3.7855)(26,3.7883)(31,3.7981)(36,3.8042)(37,3.8046)(46,3.8102)(51,3.8111)(56,3.8178)(57,3.8183)(63,3.8165)(71,3.8193)(76,3.8206)(78,3.8219)(86,3.8227)(91,3.8232)(94,3.8259)(96,3.8243)(104,3.8244)(111,3.8242)(114,3.8257)(121,3.8238)(126,3.8259)(129,3.8262)(131,3.8243)(140,3.825)(146,3.8265)(156,3.8272)(165,3.8275)(166,3.8273)(174,3.829)(176,3.8289)(184,3.8289)(189,3.8262)(191,3.8257)(199,3.8279)(203,3.8276)(210,3.8294)(212,3.8311)};
    \addlegendentry{\footnotesize{TNN TL 1\%}};

    \addplot[color=blue, mark=o, very thick]   
    coordinates {(1,3.5457)(2,3.5475)(3,3.5538)(4,3.5638)(5,3.5884)(11,3.72)(15,3.7473)(20,3.7746)(26,3.7836)(28,3.7876)(36,3.795)(44,3.7991)(47,3.7991)(53,3.7991)(57,3.7991)(61,3.7991)(66,3.7991)(71,3.7991)(76,3.7991)(81,3.7991)(87,3.7991)(91,3.7991)(96,3.7991)(101,3.7991)(106,3.7991)(111,3.7991)(116,3.7991)(121,3.7991)(126,3.7991)(131,3.7991)(136,3.7991)(141,3.7991)(146,3.7991)(151,3.7991)};
    \addlegendentry{\footnotesize{TNN w/o TL}};
    
     \addplot[color=brown, very thick]     
     coordinates {(1,3.6678)(200,3.6678)};
     \addlegendentry{\footnotesize{SNN}};
    
     \addplot[color=magenta,dashed,very thick]     
     coordinates {(1,3.4571)(200,3.4571)};
     \addlegendentry{\footnotesize{w/o NN}};

    \end{axis}
    \node[text width=3cm] at (8.5,0.5) 
    {B)};
        \draw[>=latex, thick, <->] (2,3.15) -- (2,5.25);
    \node[text width=3cm] at (5,4) 
    {0.16 bits/symbol};
    \end{tikzpicture}
    % \vspace{-1mm}
    % \caption{9dBm  16-QAM SSMF link\\ (5x50~km) - LAB B Trial.}
    \label{fig3:b} 
    \vspace{4ex}
  \end{subfigure}%% 
    \begin{subfigure}[b]{0.33\linewidth}
    \centering
 \begin{tikzpicture}[scale=0.6]
    \begin{axis} [ylabel={MI [bits/symbol]}, 
        xlabel={Training Epochs},
        grid=both,
    	xmin=1, xmax=100,
    	%xtick={-6, ..., 2},
    % 	ymin=0.00001, ymax=0.1,
    %     ymode=log,
        %log ticks with fixed point,
        legend style={legend pos=south east, legend cell align=left,fill=white, fill opacity=0.6, draw opacity=1,text opacity=1},
    	grid style={dashed}]
        ]
        \addplot[color=red, mark=square, very thick]
  coordinates {(1,3.6479)(3,3.6518)(4,3.655)(8,3.6532)(16,3.6581)(20,3.6638)(24,3.6701)(26,3.6701)(31,3.6701)(36,3.6701)(41,3.6701)(46,3.6701)(52,3.6701)(59,3.6701)(66,3.6701)(73,3.6701)(76,3.6701)(81,3.6701)(87,3.6701)(92,3.6701)(100,3.6701)(105,3.6701)(108,3.6701)(111,3.6701)(116,3.6701)(121,3.6701)(126,3.6701)(131,3.6701)(136,3.6701)(141,3.6701)(146,3.6701)(151,3.6701)(156,3.4704)(161,3.4704)(166,3.4704)(171,3.4704)(176,3.4704)(181,3.4704)(186,3.4704)(191,3.4704)(196,3.4704)(201,3.4704)};
    \addlegendentry{\footnotesize{TNN TL 100\%}};

%     \addplot[color=black, mark=star, very thick]
%   coordinates {(1,5.1465)(2,5.2206)(3,5.3163)(4,5.2761)(5,5.3176)(11,5.4004)(11,5.4004)(19,5.4428)(25,5.44)(29,5.4541)(36,5.4622)(36,5.4622)(43,5.4664)(48,5.4693)(54,5.4512)(60,5.4984)(62,5.4961)(68,5.4653)(71,5.4652)(76,5.4614)(81,5.4541)(87,5.4813)(93,5.4762)(97,5.4732)(106,5.4596)(111,5.4757)(112,5.4596)(116,5.4732)(121,5.4614)(126,5.4614)(131,5.4732)(136,5.4614)(141,5.4732)(146,5.4614)(151,5.4614)(156,5.4614)(161,5.4732)(168,5.476)(173,5.476)(176,5.4596)(182,5.476)(187,5.4596)(191,5.4596)(197,5.476)(206,5.4596)};
%     \addlegendentry{\footnotesize{TNN TL 50\%}};
    
%     \addplot[color=green, mark=star, very thick]
%   coordinates {(1,4.9821)(2,5.1167)(3,5.2042)(4,5.2114)(5,5.2707)(9,5.3024)(15,5.3462)(19,5.3497)(22,5.3908)(29,5.4129)(32,5.4108)(41,5.4126)(46,5.4509)(46,5.4509)(54,5.4496)(59,5.4257)(65,5.4371)(67,5.459)(73,5.4631)(78,5.4757)(82,5.4849)(90,5.4686)(96,5.4417)(100,5.4821)(103,5.4687)(108,5.4691)(113,5.4742)(118,5.4783)(124,5.4531)(127,5.4526)(134,5.4691)(136,Inf)(141,5.4691)(146,5.4417)(151,5.4417)(156,5.4417)(161,5.4691)(168,5.476)(173,5.476)(176,5.4596)(182,5.476)(187,5.4687)(191,5.4687)(197,5.476)(206,5.4687)};
%     \addlegendentry{\footnotesize{TNN TL 20\%}};
    
    \addplot[color=violet, mark=|, very thick]
  coordinates {(1,3.5819)(2,3.6007)(3,3.6156)(4,3.629)(5,3.6389)(10,3.6475)(16,3.6536)(18,3.654)(22,3.6568)(28,3.6528)(36,3.6543)(39,3.6545)(42,3.653)(50,3.652)(54,3.6531)(61,3.6508)(64,3.6512)(71,3.6509)(74,3.6529)(79,3.6527)(85,3.6527)(89,3.6512)(94,3.6534)(101,3.6528)(107,3.6522)(112,3.6541)(117,3.6537)(123,3.6529)(128,3.6584)(133,3.6533)(139,3.6566)(142,3.6569)(151,3.6561)(154,3.6593)(161,3.6577)(166,3.6586)(167,3.6597)(172,3.661)(178,3.6632)(185,3.6611)(191,3.6621)(195,3.6664)(199,3.6629)(201,3.6626)};
    \addlegendentry{\footnotesize{TNN TL 10\% }};

%     \addplot[color=brown, mark=star, very thick]
%   coordinates {(1,4.7491)(2,4.8843)(3,4.9269)(4,4.9976)(5,5.0499)(11,5.1502)(15,5.2111)(20,5.2567)(24,5.2771)(29,5.2948)(32,5.3011)(38,5.3057)(44,5.3049)(46,5.3036)(55,5.3694)(60,5.3296)(65,5.3508)(67,5.3685)(76,5.3631)(77,5.3709)(86,5.3799)(86,5.3799)(92,5.3503)(97,5.3934)(104,5.3822)(107,5.38)(113,5.3958)(120,5.4031)(126,5.3907)(126,5.3907)(136,5.3952)(136,5.3952)(146,5.4545)(146,5.4545)(152,5.4091)(158,5.3938)(164,5.3974)(167,5.429)(175,5.4444)(181,5.3982)(184,5.4116)(192,5.3982)(201,5.4444)};
%     \addlegendentry{\footnotesize{TNN TL 5\% }};
    
    \addplot[color=green, mark=star, very thick]
  coordinates {(1,3.4843)(2,3.5245)(3,3.5509)(4,3.5644)(5,3.5723)(11,3.582)(16,3.592)(20,3.6016)(26,3.6069)(31,3.6189)(34,3.6229)(41,3.631)(46,3.6359)(51,3.6368)(56,3.6409)(66,3.6441)(70,3.6455)(71,3.6448)(81,3.645)(86,3.648)(87,3.6486)(91,3.6464)(101,3.6476)(105,3.6498)(106,3.6494)(114,3.6475)(119,3.6486)(123,3.647)(130,3.6503)(131,3.6499)(139,3.6503)(142,3.6503)(146,3.6496)(156,3.65)(157,3.6508)(166,3.6532)(167,3.6536)(176,3.6517)(179,3.652)(186,3.6532)(187,3.6537)(193,3.6548)(196,3.6512)(204,3.6503)(211,3.6534)(216,3.653)(226,3.6533)(227,3.6537)(231,3.6521)(236,3.6521)};
    \addlegendentry{\footnotesize{TNN TL 1\%}};

    \addplot[color=blue, mark=o, very thick]   
    coordinates {(1,3.289)(2,3.2935)(3,3.31)(4,3.327)(5,3.3954)(11,3.5497)(16,3.5925)(21,3.6114)(26,3.6283)(30,3.6366)(32,3.6442)(37,3.6442)(41,3.6442)(46,3.6442)(53,3.6442)(56,3.6442)(63,3.6442)(66,3.6442)(71,3.6442)(76,3.6442)(84,3.6442)(89,3.6442)(92,3.6442)(97,3.6442)(104,3.6442)(111,3.6442)(120,3.6442)(123,3.6442)(127,3.6442)(133,3.6442)(138,3.6442)(146,3.6442)(151,3.6442)(161,3.404)(170,3.4139)(175,3.3866)(177,3.3804)(181,3.3766)(187,3.3842)(195,3.3749)(196,3.3596)(202,3.3581)};
    \addlegendentry{\footnotesize{TNN w/o TL}};
    
    \addplot[color=brown, very thick]     
     coordinates {(1,3.4917)(200,3.4917)};
     \addlegendentry{\footnotesize{SNN}};
    
     \addplot[color=magenta,dashed,very thick]     
     coordinates {(1,3.2701)(200,3.2701)};
    \addlegendentry{\footnotesize{w/o NN}};

    \end{axis}
     \node[text width=3cm] at (8.5,0.5) 
    {C)};
        \draw[>=latex, thick, <->] (2,3.15) -- (2,5.25);
    \node[text width=3cm] at (5,4) 
    {0.18 bits/symbol};
    \end{tikzpicture}
    % \vspace{-1mm}
    % \caption{3dBm  16-QAM  TWC link \\ (9x50~km) - LAB B Trial.}
    \label{fig3:c} 
    \vspace{4ex}
  \end{subfigure}%% 
\vspace{-11mm}
  \caption{\footnotesize Performance of the proposed calibration method for three experimental setups: A) 7 dBm  64-QAM  SSMF link, B) 9 dBm  16-QAM  SSMF link, and C) 3 dBm  16-QAM  TWC link. We highlight the gap between the SNN and the TNN TL by the black two-headed arrows.}
  \label{fig3} \vspace{-5mm}
\end{figure*}

Fig.~\ref{fig3} depicts the comparative study carried out over three experiments. We have considered four different cases:
i) when only linear equalization is applied (labeled as ``w/o NN''); ii) when training the NN with just experimental data (``TNN w/o TL''); iii)  when the equalizer is trained only with synthetic data (``SNN''); iv) when pre-training with synthetic data and then training the NN using TL with x\% of the experimental data used in the case ii) (``TNN TL x\%''). We studied the cases corresponding to using 100\%, 10\%, and 1\% of the original experimental dataset. Fig.~\ref{fig3} shows that some MI improvement can already be achieved when training the NN with synthetic data only (SNN). Indeed, when comparing the SNN and the reference w/o NN cases, we show a MI improvement of 0.23, 0.21, and 0.22 bits/symbol for experiments A, B, and C, respectively. Additionally, Fig.~\ref{fig3} shows also that the same performance achieved after training the NN w/o TL can be accomplished with just 1 epoch by using TNN TL and the same available experimental data (100\%). This effect is known in the machine learning literature as one-shot learning and shows how efficacious the calibration method is. In order to make the use of NN-based equalizers even more interesting, it would be desirable to reduce the amount of experimental data that is required. Thus, we studied the impact of reducing the experimental training dataset down to 10\% and 1\% of its original size. As can be observed, even with just 1\% of the original dataset ($\approx$ 2.6k symbols), the NN can still achieve the same level of max MI in a similar number of epochs as the TNN w/o TL. When using the TL, we achieved slightly higher MI values than with the TNN w/o TL, even though both were tested using the same dataset. This is in line with the known fact that TL mitigates overfitting, therefore rendering more general NN models.
%\vspace{-1mm}
\vspace{-1.5mm}
\section{Conclusions}
\vspace{-1mm}
We demonstrated an efficient method based on domain randomization and transfer learning that considerably reduces the need for experimental training resources (up to 99\%)  while still enabling the efficient equalizer's operation in realistic transmission scenarios. This is an important step for the practical realization of NN-based equalization since it provides the necessary flexibility for the NNs' adaptation to real-world scenarios.\\

\footnotesize
\linespread{0.0}
\textbf{Acknowledgements}: This work has received funding from: EU Horizon 2020 program under the Marie Sk\l{}odowska-Curie grant agreement No. 813144 (REAL-NET). SKT acknowledges the support of the EPSRC project TRANSNET (EP/R035342/1). ADE and AA acknowledge the support of the EPSRC project EEMC: EP/S016171/1, and PHOS: EP/S003436/1
\normalsize
\linespread{1.0}

\end{document}